\documentclass[apjl]{emulateapj}
\usepackage[usenames]{color}
\usepackage{hyperref}
\hypersetup{
  pdftitle = {Giant Molecular Clouds and Star Formation in the Tidal Molecular Arm of NGC~4039},
  pdfkeywords = {pdf, hyperref, bookmarks},
  pdfauthor = {Daniel Espada et al.}
}

\newcommand{\kms}{\mbox{km s$^{-1}$}}

\newcommand{\hi}{\mbox{\ion{H}{1}}}

\slugcomment{Final Draft version October 12, 2012}

\shorttitle{Giant Molecular Clouds and Star Formation in the Tidal Molecular Arm of NGC~4039}
\shortauthors{D. Espada et al.}

\begin{document}

\title{Giant Molecular Clouds and Star Formation in the Tidal Molecular Arm of NGC~4039}

\author{D. Espada\altaffilmark{1}, 
S. Komugi\altaffilmark{1,2}, 
E. Muller\altaffilmark{1}, 
K. Nakanishi\altaffilmark{1,2,3}, 
M. Saito\altaffilmark{1,2}, 
K. Tatematsu\altaffilmark{1}, 
S. Iguchi\altaffilmark{1}, 
T. Hasegawa\altaffilmark{1,2},   
N. Mizuno\altaffilmark{1,2},
D. Iono\altaffilmark{1,4}, 
S. Matsushita\altaffilmark{2,5}, A. Trejo\altaffilmark{5}, E. Chapillon\altaffilmark{5},  S. Takahashi\altaffilmark{5}, Y.N. Su\altaffilmark{5}, A. Kawamura\altaffilmark{1}, E. Akiyama\altaffilmark{1}, M. Hiramatsu\altaffilmark{1}, H. Nagai\altaffilmark{1}, R. E. Miura\altaffilmark{1}, Y. Kurono\altaffilmark{1}, T. Sawada\altaffilmark{1,2}, A. E. Higuchi\altaffilmark{1,2}, K. Tachihara\altaffilmark{1,2}, K. Saigo\altaffilmark{1}, 
T. Kamazaki\altaffilmark{1,2}}

\altaffiltext{1}{National Astronomical Observatory of Japan (NAOJ), 2-21-1 Osawa, Mitaka, Tokyo 181-8588, Japan; daniel.espada@nao.ac.jp}
\altaffiltext{2}{Joint ALMA Observatory, Alonso de C\'{o}rdova 3107, Vitacura, Santiago 763-0355, Chile}
\altaffiltext{3}{The Graduate University for Advanced Studies (SOKENDAI), Department of Astronomical Science, 2-21-1 Osawa, Mitaka, Tokyo 181-8588, Japan}
\altaffiltext{4}{Nobeyama Radio Observatory, NAOJ, Minamimaki, Minamisaku, Nagano, 384-1305, Japan}
\altaffiltext{5}{Academia Sinica, Institute of Astronomy and Astrophysics, P.O. Box 23-141, Taipei 10617, Taiwan}

\begin{abstract}

The properties of tidally induced arms provide a means to study molecular cloud formation and the subsequent star formation under environmental conditions which in principle are different from quasi stationary spiral arms. We report the properties of a newly discovered molecular gas arm of likely tidal origin at the south of NGC 4039 and the overlap region in the Antennae galaxies,  with a resolution of 1\farcs68~$\times$~0\farcs85, using the Atacama Large Millimeter/submillimeter  Array science verification CO(2--1) data. The arm extends 3.4~kpc (34\arcsec) and is characterized by widths of $\lesssim$ 200~pc (2\arcsec) and velocity widths of typically $\Delta V$~$\simeq$~10--20~\kms . About 10 clumps are strung out along this structure, most of them unresolved, with average surface densities of $\Sigma_{\rm gas}$ $\simeq$ 10--100~$M_\odot$~pc$^{-2}$, and masses of (1--8)$\times$10$^6$~$M_\odot$. These structures resemble the morphology of beads on a string, with an almost equidistant separation between the beads of about 350 pc, which may represent a characteristic separation scale for giant molecular associations. We find that the star formation efficiency at a resolution of 6\arcsec~(600~pc) is in general a factor of 10 higher than in disk galaxies and other tidal arms and bridges. This arm is linked, based on the distribution and kinematics, to the base of the western spiral arm of NGC 4039, but its morphology is different to that predicted by high-resolution simulations of the Antennae galaxies.

\end{abstract}

\keywords{galaxies: individual (Antennae galaxies) -- galaxies: ISM -- galaxies: structure -- stars: formation}

\section{Introduction}

The gas and star formation (SF) properties in arms induced by tidal interactions are believed to differ from quasi-stationary spiral arms in spiral density wave theory \citep[e.g.][]{2011EAS....52...87D}. 
However, the studies of the molecular properties in tidal arms are quite limited, even though they are important elements to shed light into our ideas of cloud fragmentation and SF process. In addition, the formation of these features contributes to the increase of star formation rate (SFR) in interacting galaxies \citep[e.g.][]{2010ApJ...720L.149T}.  
Existing studies are focused on tidal tails, where the molecular gas, as traced by CO, is only detected in a few rather disconnected clumps, and usually within \hi\ tails \citep[e.g.][]{2001ApJ...562L..43T,2004A&A...426..471L,2006AJ....132.2289W}. Surprisingly, the SF properties estimated in several collisional debris show similar values to those measured in galactic disks \citep{2001A&A...378...51B,2011A&A...533A..19B,2011arXiv1112.1922D}. 

\begin{figure}
\centering
\includegraphics[width=16.cm]{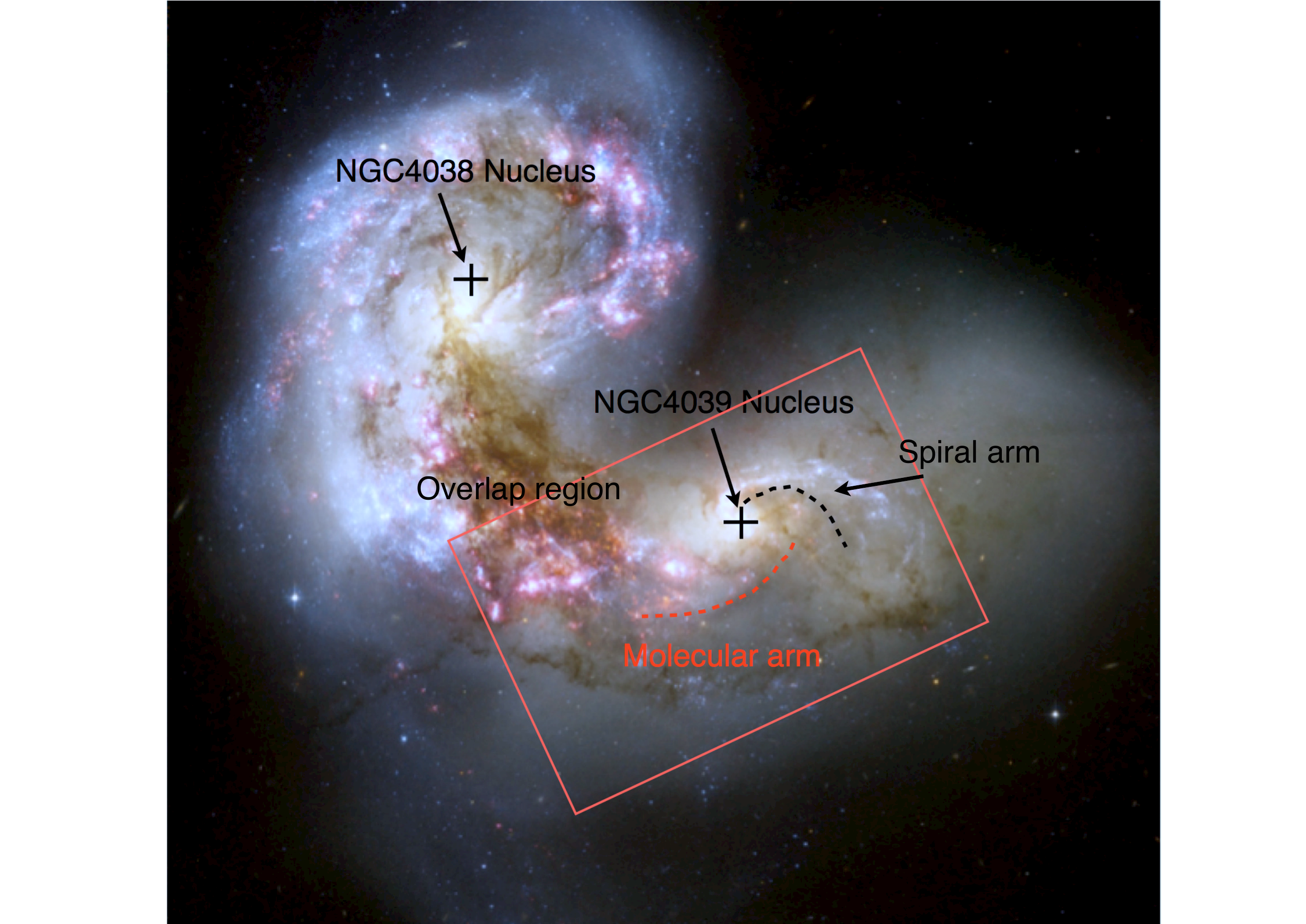}
\caption{HST Optical composite image of Antennae galaxy \citep{2010AJ....140...75W}, indicating the positions of the NGC~4038/9 nuclei, the overlap region, NGC~4039's spiral arm, and the molecular arm of tidal origin that we study in this paper. The rectangle indicates the mosaic covered by ALMA observations. \label{fig1}}
\end{figure}

One of the best studied colliding systems is the Antennae galaxy (NGC~4038/9, Figure~\ref{fig1}), partly because it is only at 22~Mpc {\bf distance} from us \citep{2008AJ....136.1482S}, and thus it is a good target to study the properties of molecular gas in tidal arms with high spatial resolution (1\arcsec\ roughly corresponds to 100~pc). The two spiral galaxies started to interact only 400--600 million years ago \citep{1993ApJ...418...82M,2010ApJ...720L.149T}, making this system one of the youngest examples of an on-going major galaxy merger. Two prominent \hi\ tidal tails extend about 70~kpc north and south (e.g. \citealt{2001AJ....122.2969H}), which have been successfully used among other information to constrain numerical simulations (e.g. \citealt{2010ApJ...720L.149T}). 

High-resolution maps ($<$ 5\arcsec) of the three lowest transitions of CO have been presented by \citet{2000ApJ...542..120W}, \citet{2012ApJ...745...65U} and \citet{2012ApJ...750..136W}  using the Owens Valley Radio Observatory (OVRO), the Plateau de Bure Interferometer (PdBI), and the Submillimeter Array (SMA).  The molecular gas as traced by CO emission is mostly distributed around the two nuclei of NGC~4038/9, a region to the west of NGC~4038, and along an overlap region between them,  and coexists with a large amount of super stellar clusters \citep[SSCs][]{1995AJ....109..960W}. 

In this Letter, we report the molecular gas properties of a tidally induced molecular arm found in NGC~4039 that had not been previously detected. We have selected this region because it is a quite narrow and elongated molecular feature that allows us to study giant molecular {\bf cloud} formation in a filamentary structure. In addition, it is not affected by projection with other components. 

\section{CO(2-1) ALMA Observations and data reduction} \label{observationReduction}

We use the Atacama Large Millimeter/submillimeter Array (ALMA) science verification CO(2--1) data of the Antennae galaxies, retrieved from the ALMA science portal\footnote{http://almascience.nao.ac.jp/almadata/sciver/AntennaeBand6/}. 
Band 6 observations with 14 antennas (11 North American and 3 East Asian antennas) were carried out in 2011 May and June, using two spectral windows of 1.875~GHz bandwidth (or $\sim$1600~\kms) and 3840 0.488 MHz channels, and with one of the spectral windows centered on the CO(2--1) line. The observations were split to cover separately the northern and southern galaxy, with 13 and 17 pointings, respectively. In this Letter, we focus on the southern mosaic, which includes the overlap region and NGC~4039. The synthesized beam is characterized by a FWHM size of 1\farcs68~$\times$~0\farcs85 and a position angle of 78$^{\rm  o}$. The covered area was approximately 1\farcm5 $\times$ 0\farcm9, along a position angle of 60$^{\rm  o}$  (Figure~\ref{fig1}). The observed UV range was 14 -- 236\,m.

Editing, calibration, and imaging were performed in a standard manner using CASA\footnote{http://casa.nrao.edu}, similar to that presented in the CASA guide\footnote{http://casaguides.nrao.edu/index.php?title=AntennaeBand7} for the CO(3--2) science verification data of Antennae galaxies\footnote{http://almascience.nao.ac.jp/almadata/sciver/AntennaeBand7/}. Due to the large complexity of the emission we restricted the cleaning to manually selected boxes where emission was apparent for each channel. 
The rms of a free-line 20~\kms\ channel is 1.2\,mJy\,beam$^{-1}$. However, this image is dynamic range limited due to the bright emission in NGC\,4039 and the overlap region, and we have estimated that the rms in each channel close to these features is $\sim$\,6\,mJy\,beam$^{-1}$. The absolute flux accuracy is estimated to be about 10\%.

Figure~\ref{fig2} shows the integrated intensity (moment 0), velocity field (moment 1), and velocity dispersion (moment 2) maps. 
To separate real emission from noise, we computed only those regions which show emission in three consecutive channels of 10~\kms\  in a data cube that was convolved to 5 pixels (0\farcs2 pixel$^{-1}$).
 The velocity range used is 1300 -- 1800~\kms. In addition, we used a clipping of 1.2\,mJy\,beam$^{-1}$ per channel for the moment 0 map, and  1.5\,mJy\,beam$^{-1}$ for the moment 1 and 2 maps. 

\section{Distribution and kinematics of the molecular arm} 
\label{sec3}

\begin{figure}
\centering
\includegraphics[width=18cm]{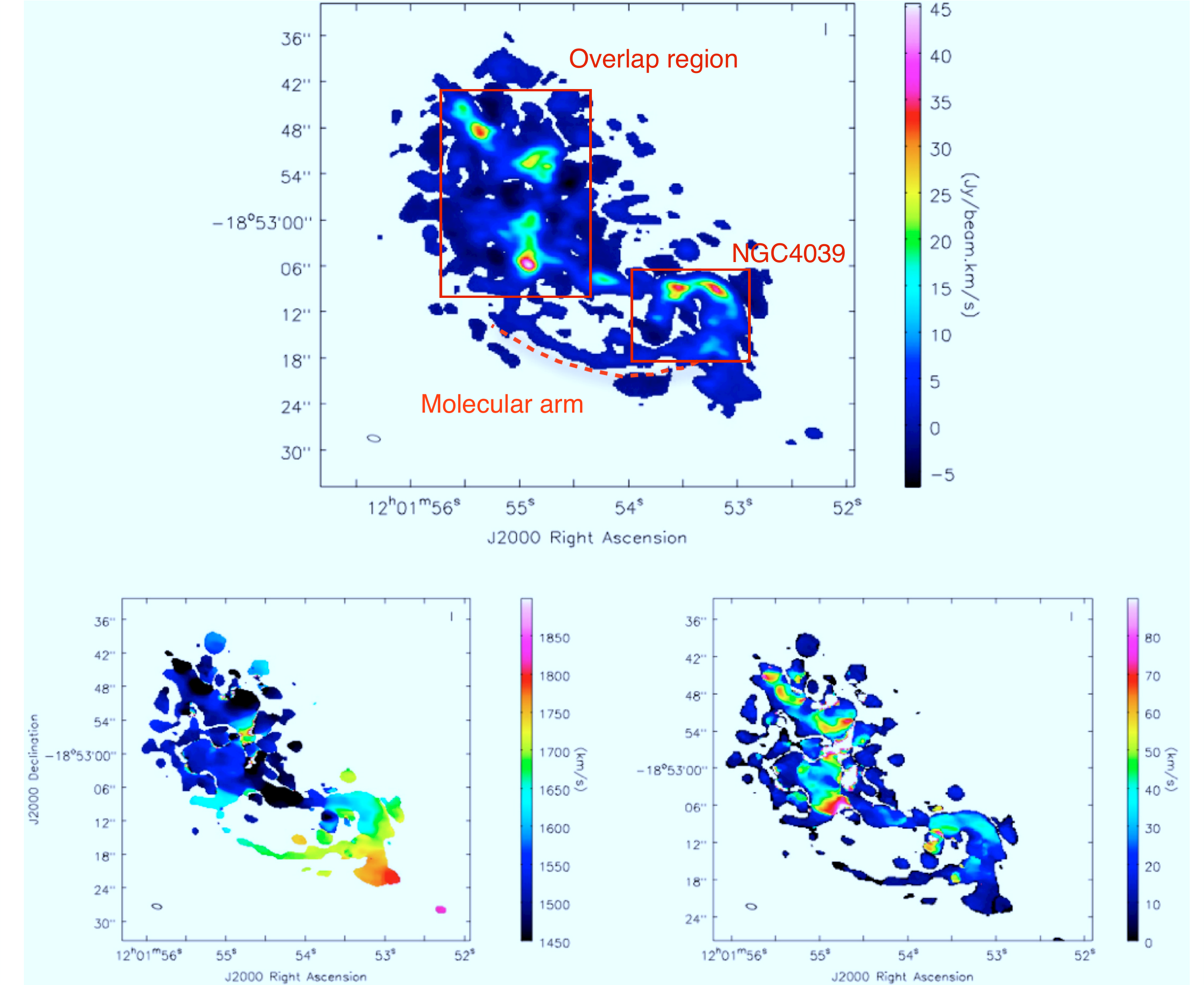}\\

\caption{ALMA CO(2--1) moment maps of the southern mosaic. {\bf (Upper panel)} The moment 0 map, integrated from 1300 to 1800~\kms . We indicate the location of the molecular arm (dashed line), the overlap region and the NGC~4039 nucleus. The boxes indicate the regions where we obtained the SF law plot in Figure~\ref{fig3}.
{\bf (Lower panel)} Velocity field (moment 1; left), from 1450 to 1900~\kms;  and the velocity dispersion map (moment 2; right), from 0 to 90~\kms. 
The beam is indicated as an ellipse in the lower left {\bf corner} of each map (HPBW of 1\farcs68 $\times$ 0\farcs85 and a PA of 78$^{\rm o}$). 
 \label{fig2}}
\end{figure}

Figure~\ref{fig2} shows that the molecular emission is particularly bright in the NGC~4039 nucleus and the overlap region between the two galaxies, where it appears concentrated in five large molecular complexes, in agreement with previous CO maps \citep{2000ApJ...542..120W,2012ApJ...745...65U,2012ApJ...750..136W}. 
A comparison with these previous maps, as well as ALMA CO(3--2) data, showed that the brightest complexes are well collocated. The total CO(2--1) flux of the entire southern mosaic map is 1460~Jy~\kms. The total flux of the map is slightly larger than previous estimates by \citet{2012ApJ...750..136W}, 1358~Jy~\kms. By comparing with single dish observations, \citeauthor{2012ApJ...750..136W} infer that flux loss in their interferometric observations is $\sim$12\%, which is comparable to absolute total flux uncertainties.

The CO(2--1) map (Figure~\ref{fig2}) shows a molecular arm in the southern region close to NGC~4039 and the overlap region. The arm extends 34\arcsec\ end to end, or 3.4~kpc. Some regions of this feature are slightly resolved in the N--S direction, but the width is typically unresolved at $\sim$200~pc. The CO(2--1) flux of this feature is 60~Jy~\kms, a 4\% of the total flux in the ALMA CO(2--1) southern mosaic. This component is also partially seen in the ALMA CO(3--2) maps, although with a considerably lower signal-to-noise ratio. This may be due partly to low excitation, especially in regions further from NGC~4039 and the overlap region (Figure~\ref{fig1}).

The velocity of this component ranges from 1600 to 1750~\kms, but the velocity dispersion is typically $\sim$10--20~\kms\ (Figure~\ref{fig2}, lower left panel). This is slightly higher than the typical velocity dispersion of 5--10~\kms\ in the molecular gas of face-on galaxy disks \citep[e.g.][]{1997A&A...326..554C}. The overlap region shows much larger velocity gradients and velocity dispersions, up to $\sim$70~\kms, in agreement with \citet{2012ApJ...745...65U} and \citet{2012ApJ...750..136W}.
Figure~\ref{fig3} (upper panel) shows that the molecular arm's velocity centroids vary continuously and slowly within the velocity range 1650 -- 1750~\kms. This continuity is kept along the spiral arm of NGC~4039, but not in the vicinity of the brightest giant molecular gas complex in the overlap region, where several spectral components are found in projection.

The molecular gas surface densities were derived using $\Sigma_{\rm H2}$~($M_\odot$~pc$^{-2}$ )~= 156 $\times$~($X$/3.0~$\times$ 10$^{20}$)~S$_{\rm CO(2-1)}$$\Delta V$~[Jy~\kms~arcsec$^{-2}$], where the S$_{\rm CO(2-1)}$$\Delta V$ term is the CO(2--1) integrated flux density. We adopted an intensity CO(2--1) to CO(1--0) ratio $I(2--1)/I(1--0)$ = 1.0 and a CO-to-H$_2$ conversion factor $X$~=~3.0~$\times$~10$^{20}$~cm$^{-2}$ (K~\kms )$^{-1}$ \citep{2012ApJ...750..136W}.  We did not correct by 1.36 to account for He. 
Along the molecular arm, the gas surface densities of the brightest components are typically $\sim$ 10--100~$M_\odot$ pc$^{-2}$. Note that in the overlap region they are as high as 1000~$M_\odot$ pc$^{-2}$. 

At least 10 clumps are seen along this structure as shown in the lower panel of Figure~\ref{fig3}. 
The masses of these clumps span from (1--8)~$\times$~10$^6$~$M_\odot$, with the most massive ones being closer to NGC~4039.
Using also CO(2--1) observations, \citet{2012ApJ...750..136W} identified a distinct break in the mass function of molecular clouds at $\sim$ 3~$\times$~10$^{6}$~$M_\odot$, with higher values corresponding to actively forming regions. The clumps in the molecular arm are therefore in between these two mass populations.
For a cloud size scale of $\sim$100 pc, using a typical velocity dispersion of 10~\kms, and a geometric factor $\alpha_g$ for the density profile of 5/3 as in \citet{2012ApJ...750..136W}, the virial mass is 7$\times$10$^6$~$M_\odot$. This suggests that while the molecular clumps close to NGC~4039 are compatible with structures that are bound gravitationally, we cannot discard that they are not fully bound further away. Higher resolution observations to resolve these complexes are necessary to set stronger constraints. 

The separation between these complexes (peak to peak average value) is found to be almost equidistant at about 350~pc, assuming that the effect of inclination is negligible. It is remarkable the resemblance to classical "beads on a string" \citep[e.g.,][]{1983MNRAS.203...31E} along the molecular arm.
The characteristic length ($\lambda$) to filament diameter ($D$) ratio, $\lambda$/$D$,
is $\lambda$/$D$ $\simeq$ 2, assuming $\lambda$ = 350~pc and $D$~=~200~pc. 

Although usually applied to different scales than those presented here, a number of models have been created to describe fragmentation of filamentary structures. If fragmentation is caused by self-gravity the distance between adjacent clumps is of the order of the Jeans length. The typical Jeans length depends on gas pressure (both thermal and turbulence), magnetic field and rotation. The expected $\lambda$/$D$ ratio obtained in these models, $\lambda$/$D$, is typically 2--4, with smaller ratios found for faster rotation and/or stronger magnetic fields  \citep[e.g.,][]{1993PASJ...45..551N,1993ApJ...404L..83H}, in agreement with the CO(2--1) observations.  

The filamentary structure of the molecular arm is unique to probe the characteristic separation between giant molecular associations (GMAs), and allows us to study the typical size scales for fragmentation occurring between parsec and kiloparsec scales.
Observations of Galactic filamentary clouds show characteristic separations of about three times their diameter \citep[a few parsec scale; e.g.,][]{1979ApJS...41...87S}.
The separations of GMAs in nearby galaxies and our Galaxy are characterized by a large scatter, with separations of a few 100~pc to 1~kpc \citep[e.g.,][]{1993prpl.conf..125B,2006ApJ...638..191K}. These are located within larger scale complexes of SF regions separated {\bf on} average a few kiloparsecs and typical diameters of 0.5--1.0~kpc, in a configuration that also  resembles beads on string morphology \citep[e.g.][]{1983MNRAS.203...31E,2006ApJ...642..158E,2010AJ....139.1212S,2011ApJ...731...93M}. 

The general properties of the molecular arm are comparable to that of spiral arms in non-interacting galaxies. However, the width seems smaller than the typical width in spiral arms, usually in the range of 400--1500\,parsec based on measurements using H$\alpha$ maps \citep[e.g.,][]{1982ApJ...253..101K}. 
Although the inclination of this structure is rather uncertain, we find it unlikely that the observed molecular arm can be reproduced with a single pitch angle, which may indicate that it has been created or affected by tidal interactions.

\section{Star Formation Law and Stellar Clusters}
\label{sec4}

In order to investigate the SF properties in the molecular arm, we derived the SFR surface densities ($\Sigma_{\rm SFR}$) from available FUV and 24~$\mu$m maps.
We used FUV data from the Nearby Galaxy Survey \citep{2007ApJS..173..185G}. The FUV band covers the wavelength range 1350--1750 \AA, and provides a resolution of 4\farcs0. We corrected the FUV maps for foreground extinction using $A_{\rm FUV}$~=~8.24$\times$ $E(B--V)$ \citep{2007ApJS..173..293W}, where $E(B--V)$ = 0.046~mag as estimated by \citet{schlegel98}.   
We use the 24~$\mu$m map presented in \citet{2010MNRAS.401.1839Z}, which was obtained with the MIPS \citep{2004ApJS..154...25R} on board the Spitzer Space Telescope, and was corrected for background, matching of individual data frames, cosmic-ray removal, flat-fielding and mosaicking using MOPEX. The FWHM of the MIPS maps point-spread function (PSF) at 24~$\mu$m is 6\arcsec.  The FUV map was convolved to the same spatial resolution of 6\arcsec.
The $\Sigma_{\rm SFR}$ was calculated following \citet{2008AJ....136.2782L}: $\Sigma_{\rm SFR}$~[$M\sun$ yr$^{-1}$ kpc$^{-2}$ ] = 3.2~$\times$~10$^{-3}$ $I_{24}$ [MJy str$^{-1}$ ] + 8.1 $\times$ 10$^{-2}$ $I_{\rm FUV}$ [MJy str$^{-1}$]. The sensitivity limit of the $\Sigma_{\rm SFR}$ is 1 $\times$ 10$^{-4}$ $M\sun$ yr$^{-1}$~kpc$^{-2}$.

We also convolved the molecular gas surface densities to 6\arcsec\, and converted the molecular gas and SFR surface density maps to a pixel size of 3\arcsec . 
For the SFR map, a 3$\sigma$ threshold was used, or 0.018~$M_\odot$ yr$^{-1}$.  
Figure~\ref{fig4} shows the pixel-to-pixel SF law plot for different regions in the southern mosaic.  
The average star formation efficiency (SFE), the SFR per unit of gas,  for the molecular arm is log(SFE [yr$^{-1}$]) = --8.2 $\pm$ 0.3, similar to that of the NGC~4039 region log(SFE [yr$^{-1}$]) = --8.8 $\pm$ 0.3, and both are a factor of 10 lower than in the overlap region. These are still larger than the average SFE for normal galaxies of log(SFE [yr$^{-1}$]) = -9.3 \citep{2005PASJ...57..733K,2008AJ....136.2846B}. 

Note that for the molecular arm we used only regions in the SFR map that are more than 11\arcsec\ away from the brightest points in the overlap region and NGC~4039, in order to avoid confusion with the airy ring at $\sim$8\arcsec\ in the 24~$\mu$m PSF.
The molecular arm region pixel closest to the overlap peak (611~MJy~str$^{-1}$, with background removed) is about
13\arcsec\ away from the overlap peak.  The 24~$\mu$m PSF is about 1\% here, or about 6~MJy~str$^{-1}$.
The actual 24~$\mu$m flux density at this region is $\sim$24~MJy~str$^{-1}$, so the PSF contamination in this case is about 25\%, or about 0.1~dex on the SF law plot.  Considering the rms dispersion of the SFE (0.27--0.46~dex), this can be considered negligible.

A large number of bright young star clusters ($\sim$700 identified) exist in the Antennae galaxies \citep{1995AJ....109..960W,1999AJ....118.1551W}. However, it is remarkable that 
the SSCs number density is quite low along the molecular arm.  
  We can only identify five candidates of SSCs within 5\arcsec\ of the molecular arm (see Figure~\ref{fig3}). 
Based on $V-I$ colors and the color-age diagram presented in \citet{2010AJ....140...75W},  we could only constrain the age for SSC 33 \citep{1995AJ....109..960W}, $<$10~Myr. Based on the relative location and young age of SSC 33, we suggest that it is likely associated with some of the more massive molecular gas complexes in the molecular arm. 

\section{Discussion: Origin of the Molecular Arm}
 \label{conclusion}

\begin{figure}
\centering
\includegraphics[width=11.cm]{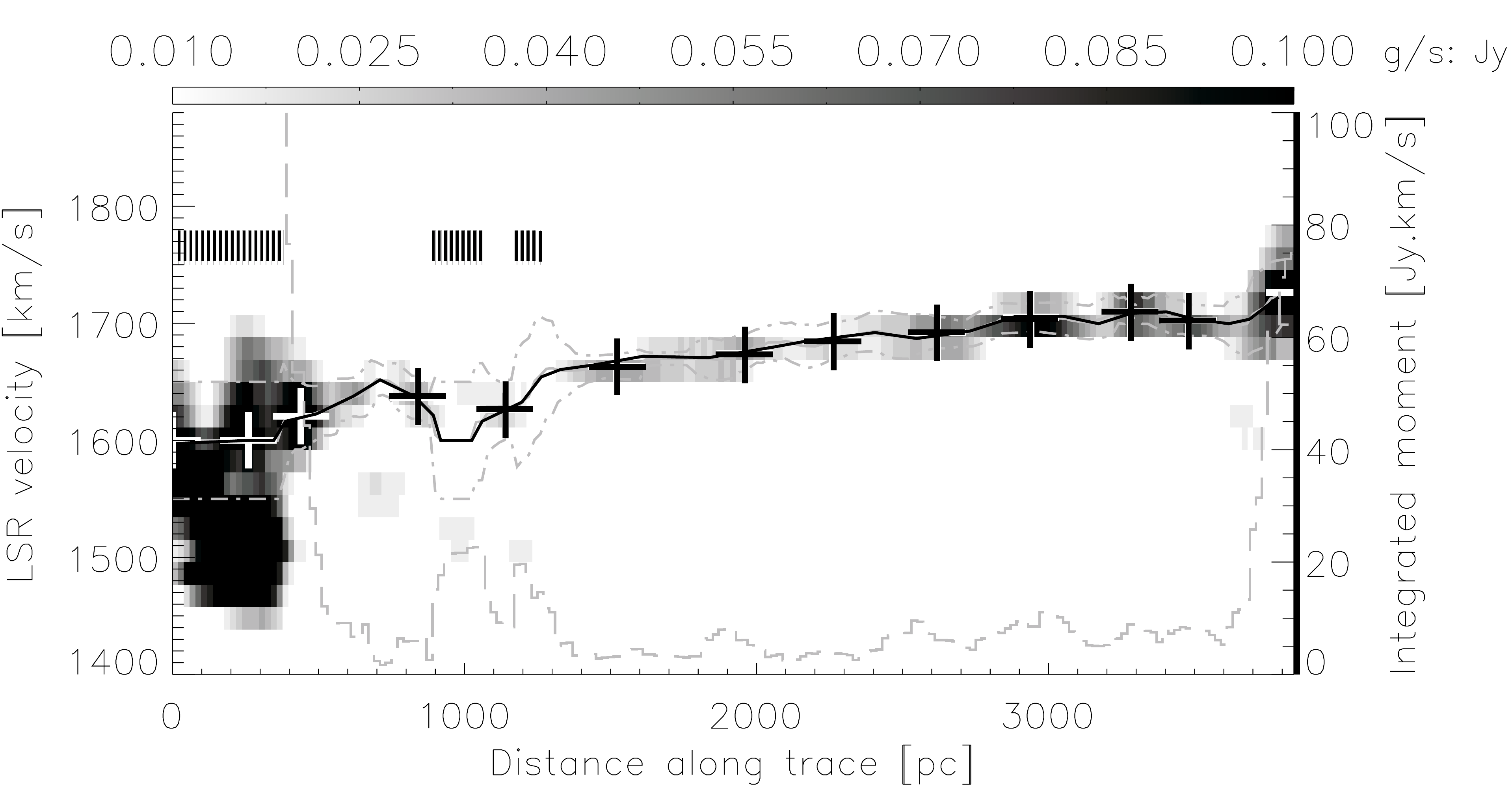}
\includegraphics[width=12.cm]{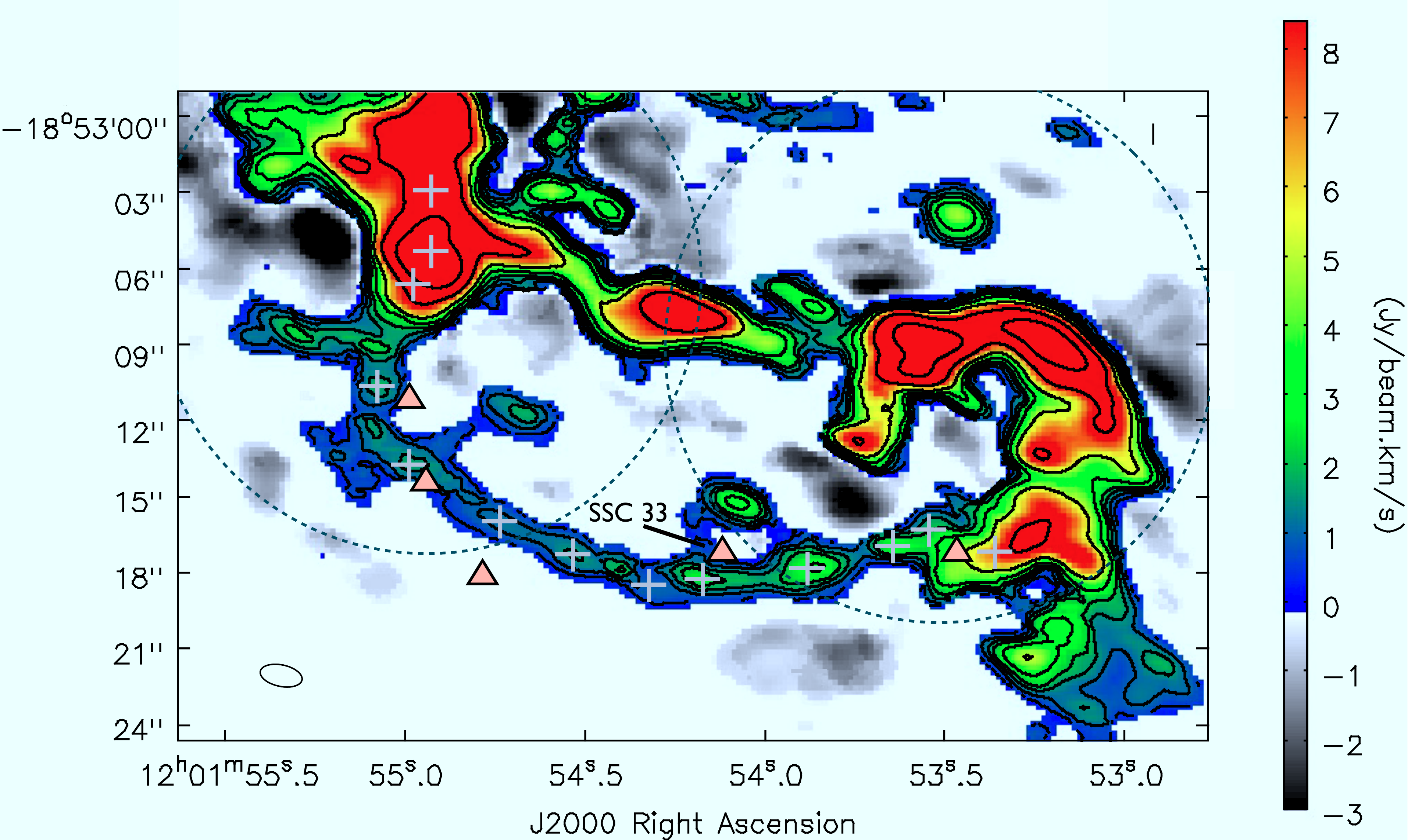}
\caption{{\bf (Upper panel)} Velocity versus distance along the trace linking the points of higher peak temperature (see lower panel). 
The solid line indicates the centroid velocity of the fit, and the dashed-dotted lines the $\pm$ 1/2 dispersion. The grey scale indicate peak intensity. The integrated flux along the trace is shown as a dashed profile. The vertical lines indicate profiles where {\bf the} fit did not converge, given initial conditions and constraints to fit. 
{\bf (Lower panel)} Integrated intensity map, emphasizing the tidal molecular arm. The contours are 0.5, 1, 1.5, 2, 3, 5, 10, 22~Jy~beam$^{-1}$~\kms. The plus signs show high local peak temperature as in the upper panel.  The (dashed line) circles show the region excluded in the SF law plot  for the molecular arm region (Section\,4 and Figure~\ref{fig4}). The triangle signs show the location of super stellar clusters along the molecular arm (see Section\,4). We indicate SSC 33, for which a reasonable accurate age could be estimated.
 \label{fig3}}
\end{figure}

\begin{figure}
\centering
\includegraphics[width=10.cm]{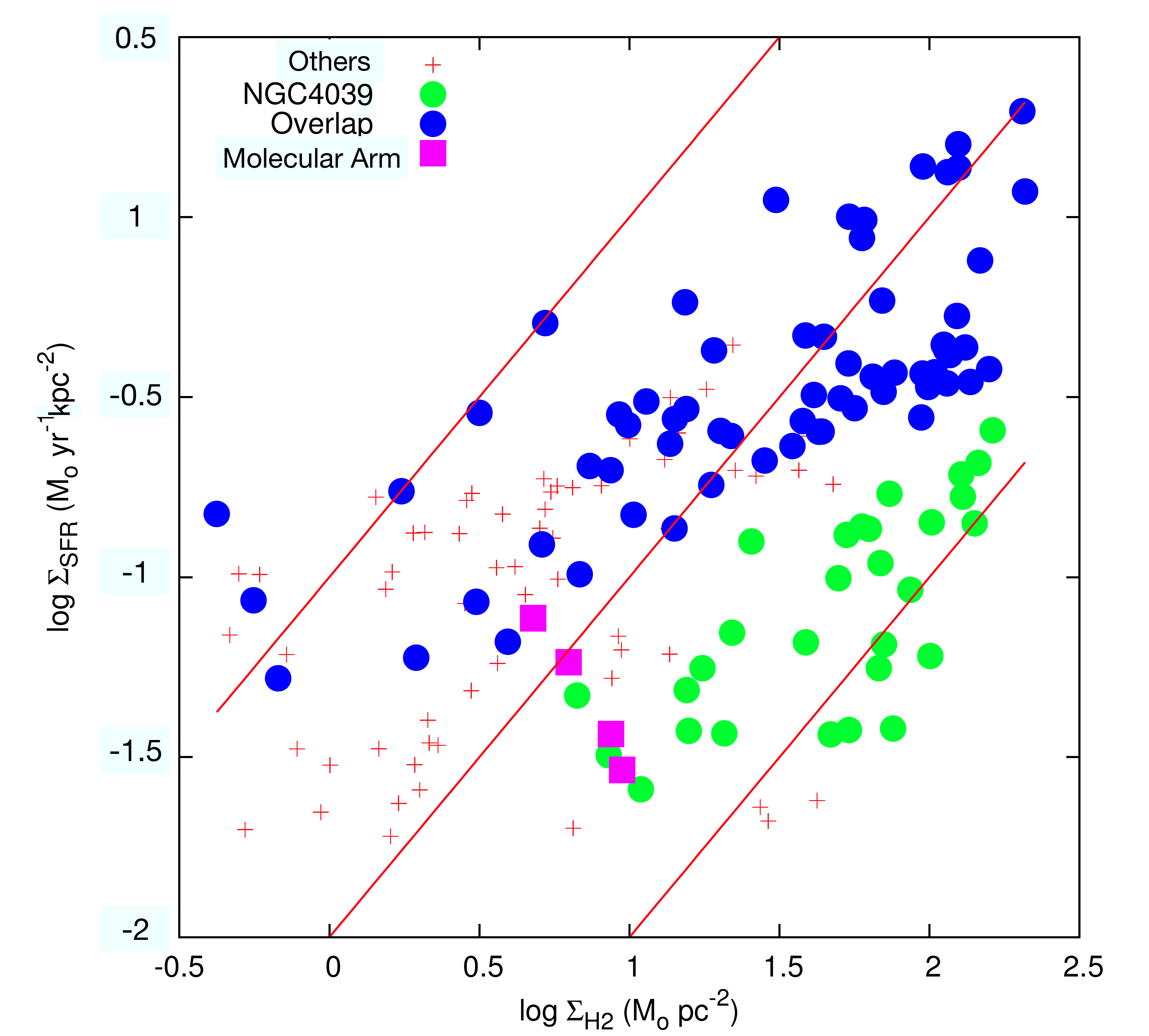}
\caption{ Star Formation Rate (SFR) versus H$_2$ surface densities (Schmidt-Kennicutt law, see Section\,\ref{sec3} and \ref{sec4}), emphasizing the  molecular arm (squares), NGC4039 (green circles), and the overlap region (blue circles). Data points for other regions in the southern mosaic are shown with plus signs. The rectangles used for each region are indicated in Figure\,\ref{fig2}. The lines indicate constant SFE, from top to bottom, at 10$^{-7}$, 10$^{-8}$, and 10$^{-9}$~yr$^{-1}$. Each data point corresponds to a pixel of 3\arcsec\  in a 6\arcsec\ resolution map. 
}
\label{fig4}
\end{figure}

  That the molecular arm is a single entity  linked to the spiral arm of NGC~4039 is proven by the smooth and continuous velocity change along its different components in the diagram shown in Figure~\ref{fig3}. 
Due to the link in terms of distribution and kinematics with the spiral arm of NGC~4039, 
we believe that the most plausible explanation is that the molecular arm is of tidal origin,  presumably caused by the recent passage of the northern galaxy, NGC~4038. 
\citet{2010AJ....140...75W} identified in Hubble Space Telescope (HST) images the molecular arm as a dark, relatively circular dust cloud (found near the so-called regions "Outer 3" and "Outer 6" in Figures 20(c) and (g), and suggested that this material may be falling back into the galaxy. 

If the stripped material contained both atomic gas (\ion{H}{1}) as well as molecular gas, then we would expect that \ion{H}{1} is also extended in this region, in disagreement with the observations. The \ion{H}{1} integrated density distribution \citep{2001AJ....122.2969H} does not show any clear correlation with the molecular arm, although the available \ion{H}{1} map's resolution is a factor of 10 worse than that of the CO(2--1) map. 
In the region of the molecular arm, the \ion{H}{1} flux density is $S_{HI}$ $<$ 0.03\,Jy\,beam$^{-1}$\,km\,s$^{-1}$, or \ion{H}{1} surface density of 1\,$M_\odot$\,pc$^{-2}$, following for example equation in \cite{2011ApJ...736...20E}.  This suggests that mostly molecular gas material was dragged.
If not of tidal origin, another explanation might be a bubble created by massive SF. However, this is unlikely due to the necessary energy output,  $E$ $\sim$ 10$^{51}$\,J following a similar method as in \citealt{2006ApJ...636..685S}, which would require $\sim$ 10$^7$ supernovae explosions.


Next we compare with available high resolution numerical simulations of the Antennae galaxies.
\citet{2010ApJ...715L..88K} recently presented  simulations with resolutions of $\sim$70~pc, using radiative cooling, SF, feedback from supernovae, and a refined merger orbit. These simulations reproduce in general the observed gas morphology and kinematics, as well as the SFR. 
However, note that the predicted projected gas surface density map (their Figure~5.3) shows significantly more gas in the southern disk than in the \ion{H}{1} maps, and \citeauthor{2010ApJ...715L..88K} ascribe it to most of the gas being in molecular rather than in atomic phase. The ALMA maps show such missing molecular gas, but the morphology is different. 
We notice that there is a filamentary structure close to the east of the overlap region and crossing it, as also seen in their young star distribution plot, and presumably extending from the northern \hi\ tidal tail to the eastern side of NGC~4039. This may indicate that the molecular arm is linked to the base of the northern \hi\ tidal tail. A deeper study with further numerical simulations would be needed to match the molecular gas morphology found in the observations. 

ALMA allows us to tackle the origin and properties of such weaker yet now detectable molecular gas components, which should be widely present in interacting systems. Data from other more compact configurations of ALMA, including zero and short spacings provided by the Atacama Compact Array \citep{2009PASJ...61....1I}, are essential to reveal the distribution and kinematic properties of the gas in such features.

\acknowledgments{We thank the referee for useful comments and
suggestions. This Letter makes use of ADS/JAO.ALMA\#2011.0.00003.SV.  ALMA is a partnership of ESO (representing its member states), NSF (USA) and NINS (Japan), together with NRC (Canada) and NSC and ASIAA (Taiwan), in cooperation with the Republic of Chile. The Joint ALMA Observatory is operated by ESO, AUI/NRAO and NAOJ. 
 KT is financially supported by the Grants-in-Aid for the Scientific Research
by the Ministry of Education, Science, Sports and Culture (No.~23540277).
GALEX is operated for NASA by the California Institute of Technology under NASA contract NAS5-98034. The Spitzer Space Telescope is operated by the Jet Propulsion Laboratory, California Institute of Technology, under contract with the National Aeronautics and Space Administration. }

{\it Facilities:} \facility{ALMA, GALEX, Spitzer}

\end{document}